\documentstyle[preprint,eqsecnum,aps]{revtex}
\begin{document}
\draft
\preprint{hep-lat/9306002,~~CTP\#2148}
\title{CORRELATION FUNCTIONS OF HADRON CURRENTS \\
        IN THE QCD VACUUM CALCULATED IN LATTICE QCD\footnotemark[1]}

\footnotetext[1]{This work is supported in part by funds
provided by the U. S. Department of Energy (D.O.E.) under contracts
\#DE-AC02-76ER03069 and \#DE-FG06-88ER40427, and the National Science
Foundation under grant \#PHY~88-17296.}

\author{M.-C. Chu}
\address{Kellogg Radiation Laboratory\\
         California Institute of Technology, 106--38\\
         Pasadena, California 91125}
\author{J.~M.~Grandy}
\address{T--8 Group, MS B--285\\
         Los Alamos National Laboratory\\
         Los Alamos, New Mexico 87545}
\author{S.~Huang}
\address{Department of Physics, FM--15\\
         University of Washington\\
         Seattle, Washington 98195}
\author{J.~W.~Negele}
\address{Center for Theoretical Physics\\
         Laboratory for Nuclear Science and Department of Physics\\
         Massachusetts Institute of Technology\\
         Cambridge, Massachusetts 02139}

\date{\today}
\maketitle

\begin{abstract}
Point-to-point vacuum correlation functions for spatially separated
hadron currents are calculated in quenched lattice QCD on a
$16^3\times 24$ lattice with $6/g^2=5.7$. The lattice data are analyzed
in terms of dispersion relations, which enable us to extract physical
information from small distances where asymptotic freedom is
apparent to large distances where the hadronic resonances
dominate. In the pseudoscalar, vector, and axial vector channels where
experimental data or phenomenological information
are available, semi-quantitative agreement is obtained.
In the nucleon and delta channels, where no experimental data exist,
our lattice data complement experiments.
Comparison with approximations based on sum rules and interacting instantons
are made, and technical details of the lattice calculation are described.
\end{abstract}

\widetext

\section{INTRODUCTION}

The structure of the QCD vacuum and of hadrons poses an exceedingly rich and
complicated many-body problem.  Hence, as with other many-body systems, it is
instructive to focus one's attention on appropriately selected ground state
correlation functions, to calculate their properties quantitatively, and to
understand their behavior physically.

The correlation functions we address in this work are the space-like separated
correlation functions of hadron currents in the QCD vacuum which have recently
been discussed in an extensive review by Shuryak \cite{shuryak}.
For example, in the case of a meson current
$J(x) \equiv \bar{q}(x) \Gamma q(x)$, we consider the correlation
function $\langle \Omega | TJ(x) \bar{J}(0) |\Omega\rangle$
where $x$ is space-like and for
simplicity may be taken to be purely spatial so that the two currents are at
equal time and $ | \Omega \rangle $ denotes the interacting ground state.
This correlation function has a number of appealing features.  In
many channels, it has been determined phenomenologically by using dispersion
relations to relate it to $e^+e^-$ hadron production and $\tau$-decay
experimental data.  Because it is defined at equal time (or Euclidean time),
it may be calculated on the lattice and in the interacting instanton
approximation as well as by using sum rules.  It complements bound state
hadron properties in the same way scattering phase shifts provide information
about the nucleon-nucleon force complementary to that provided by the
properties of the deuteron.  Just as nucleon-nucleon scattering allows one
to explore the spin-spin, spin-orbit and tensor components of the nuclear force
at different spatial separation in much more detail than deuteron observables
which reflect the composite effect of all channels and ranges, so also the
interaction or ``scattering'' of virtual quarks and antiquarks from meson
sources at different spatial separations allows one to obtain much more
detailed information about quark interactions for different channels and
spatial separations than the composite effects reflected in hadron bound
states.

Much of the richness of the study of these correlation functions derives from
the different physics involved at different spatial separations.  For
convenience, we will consider the ratio of the physical correlation functions
to those of massless non-interacting quarks, which by dimensional
considerations must fall as $x^{-6}$.  By asymptotic freedom, at extremely
short distances the interactions between quarks must become negligible, and
the ratio approaches 1.  For slightly larger distances, where interactions are
small but non-negligible, one should be able to use the leading terms in the
Wilson operator product expansion to describe the deviation from unity.  In
the absence of separate, exact evaluation of the relevant operators, one must
use the factorization approximations which have been developed in connection
with sum rule techniques, and we will see below that these approximations are
successful in some channels and fail in others.  At still larger distances,
the full complexity of non-perturbative QCD comes into play, and one may use
this region to test and refine QCD motivated models such as the interacting
instanton approximation.
For example, a dominant feature of instanton models is
the 't~Hooft effective interaction which couples left- and right-handed
quarks and thus in leading order contributes only to the scalar and
pseudoscalar channels and with opposite sign.  Finally, at very large
separation, the decay of the correlation functions is governed by the
lightest hadron mass in the relevant channel.

As a result of this diverse range of physics at different spatial separations,
it is clear that definitive lattice calculations of correlation functions
would provide an exceedingly useful supplement to accessible experimental data
in allowing one to quantitatively explore and improve approximations based on
the operator produce expansion, sum rules, and interacting instantons.  We
view the exploratory calculation described in this work as a successful first
step in this direction.  Although the lattice size, lattice spacing,
statistics, extrapolation to the physical pion mass,
and quenched approximation limit our present accuracy, the
semi-quantitative agreement we obtain in channels for which experimental data
exist and the lack of major pathologies in statistical noise or extrapolation
to the chiral limit clearly indicate the potential for lattice calculation of
these correlation functions.  In addition to the meson correlation functions
discussed above, we will also address analogous correlation functions for
baryon currents.

The outline for this paper is as follows.  In Section~II we define the hadron
currents we use, present the results for correlation functions of these
currents for free quarks on the lattice, and describe a useful
phenomenological parameterization in terms of resonance and continuum
contributions. The lattice calculation is described in Section~III, including
the treatment of lattice anisotropies, corrections for images, extrapolation in
$\kappa$, and determination of parameters of the phenomenological fits.
Section~IV presents the results.
For each meson and baryon channel, the lattice
correlation functions extrapolated to the physical pion mass are presented
and, where possible, compared with correlation functions extracted
from experimental data and approximations based on sum rules and instanton
approximations.  In addition, masses, coupling constants, and thresholds
extracted from phenomenological fits to the lattice results are also presented
and discussed.  Our conclusions are discussed in Section~V.

\section{CORRELATION FUNCTIONS }

The two-point function for a generic current $J$ is defined as
the vacuum expectation value of the time-ordered product
\begin{equation}
\langle\Omega|TJ(x)\bar{J}(0)|\Omega\rangle\equiv R(x) \, .
\end{equation}
For a local field theory, the two-point function in momentum space can be
uniquely characterized up to a polynomial by
its absorptive part through a dispersion relation
\begin{equation}
\int d^4x e^{iqx}\langle\Omega|TJ(x)\bar{J}(0)|\Omega\rangle
=K(q)\int ds{f(s)\over s-q^2-i\epsilon} + P(q^2) \, ,
\label{dispersion}
\end{equation}
where $K(q)$ is a kinematic factor which only depends on the
quantum numbers of $J$, and $P(q^2)$ is a finite-order polynomial
of $q^2$.

  In the case of free massless quarks, without the gauge interaction,
the spectral density function $f(s)$ can be calculated easily.
For a mesonic channel with current $J$ composed of quark and antiquark
operators and Dirac and flavor matrices  tabulated in Table~\ref{table1},
the corresponding $f(s)$ is simply given by the imaginary part of
the fermion bubble graph in the same channel. Up to an overall
sign, $f(s)=3s/8\pi^2$ and $K(q)=1$ for scalar and pseudoscalar
channels.  Similarly $f(s)=1/4\pi^2$ and $K(q)=q^2g_{\mu\nu}-q_\mu q_\nu$
for vector and axial vector channels. For convenience, we contract
indices $\mu$ and $\nu$ in vector and axial channels to obtain
direction independent correlators.

  To calculate free baryonic correlators with massless quarks, it
is more convenient to work in coordinate space, rather than in
momentum space. Otherwise, we would have to calculate the imaginary
parts of two-loop graphs. In coordinate space, a massless
quark propagator has the following simple form
\begin{equation}
\langle 0|Tq(x)\bar{q}(0)|0 \rangle
={i\over 2\pi^2}{x_\mu\gamma^\mu\over x^4}\, .
\end{equation}
For the nucleon and delta currents defined in Table~\ref{table1},
it is easy to verify the following:
\begin{equation}
\langle 0|TJ^N(x)\bar{J}^N(0)|0\rangle =i{24\over \pi^6}
{x_\mu\gamma^\mu\over x^{10}} \, ,
\end{equation}
\begin{equation}
\langle 0|TJ^\Delta_\mu(x)\bar{J}^\Delta_\mu(0)|0\rangle
=-i{18\over \pi^6}{x_\mu\gamma^\mu\over x^{10}} \, .
\end{equation}
If quarks are given small masses, there will be terms proportional
to the identity in Dirac space with coefficients linear in
quark masses. In order to have stable and quark mass insensitive
free correlators, which will be used as the normalizations for the
interacting correlators, we multiply these baryonic correlators by
a factor of $x_\mu\gamma^\mu$ and then take the trace in the Dirac
indices. Finally, a Fourier transform is necessary to obtain the
spectral density functions.

In order to divide out the lattice artifacts at small distances,
the corresponding non-interacting lattice correlators have been used
for normalization.
A small quark mass has to be introduced to render the lattice quark
propagator well-defined. It can be verified that a finite quark mass
only affects the propagator at distances larger than the Compton
wavelength. For a quark mass of the order of 50 MeV, significant
deviation from the massless continuum quark propagator only occurs
at distances beyond 4 fm. At distances less than 2 fm, the
deviation is at most a few percent, as discussed below.

  Once the gauge interaction is turned on, we do not in general know
how to calculate the spectral density functions. However, we do
know their qualitative behavior,
based on experimental information and general properties
of local field theories. So the natural strategy is to
parameterize the spectral density functions phenomenologically and
then determine the parameters by fitting them to lattice results.
For the present application, it is adequate to use the following
parameterization
\begin{equation}
f(s)=\lambda^2\delta(s-M^2)+f_c(s)\theta(s-s_0) \, ,
\label{spectral}
\end{equation}
where $M$ is the bound state mass, $\lambda$ denotes the coupling
of the current to the bound state and $s_0$ is the threshold for the
onset of a continuum contribution $f_c(s)$. Note that in this
parameterization, sharp resonances are treated as pole terms, while
broad and overlapping resonances are treated as a continuum.
The functional form of $f_c(s)$ is in general very complicated.
Due to asymptotic freedom, we expect that $f_c(s)$ approaches the
free result $f_0(s)$ for sufficiently large $s$, and for intermediate
$s$ the non-interacting result could be corrected perturbatively.
However, since the
threshold $\sqrt{s_0}$ is around 1.5 to 2.0 GeV, which is still in the
non-perturbative region, there is no analytical means to accurately
describe all detailed behavior near threshold. Hence, we will
parameterize $f_c(s)$ by the functional form of the free result
$f_0(s)$, listed also in Table~\ref{table1}, throughout
the whole continuum region, and treat the threshold $s_0$ as a
phenomenological parameter which produces the correct integrated
strength for the low $s$ non-perturbative region of the spectral
function.  Note that as a result, the value of $s_0$ need not correspond
precisely to the threshold for the first excited state found in the
particle data tables.

A sketch of a generic spectral function and its representation by the
parameterization of eq.~(\ref{spectral}) with $f_0(s)$ are shown
in Fig.~\protect\ref{sketch}(a) by the light and heavy curves respectively.
The single isolated resonance is represented by a delta-function at $M^2$.
We have chosen the case of a scalar or pseudoscalar channel, where the
non-interacting continuum spectral function grows linearly in $s$.
One observes that the linear curve  $f_0(s)$ joins the continuum
smoothly at large $s$ and with the value of $s_0$ shown, includes
roughly the same strength as the full continuum curve at low $s$.

The same kind of approximation is
widely used in QCD sum rule calculations. It should be emphasized
that using $f_0(s)$ to approximate $f_c(s)$ is consistent with
the lattice approximation, since the resonance parameters and
continuum strength near threshold are determined by physics well
below the lattice cutoff $p^2 = (\pi/a)^2 \approx 9$ GeV$^2$.
This is in contrast with the operator product
expansion based QCD sum rule calculations, in which
there may not be an appropriate region to match
the theoretical calculations and the phenomenological results.

Given the spectral density described above, Eq.~(\ref{spectral}),
an inverse Fourier transform of
Eq.~(\ref{dispersion}) defines the phenomenological correlators
in coordinate space as a function of $M$, $\lambda$ and $s_0$.
Note that the
polynomial $P(q^2)$ only contributes at the point $x=0$ and can be
ignored at finite $x$. In doing the Fourier transform the following
two integrals are used
\begin{equation}
\int {d^4q\over (2\pi)^4} e^{-iqx} {i\over q^2-s}
  ={\sqrt{s}\over 4\pi^2 x} K_1(\sqrt{s} x)
\end{equation}
and
\begin{equation}
\int {d^4q\over (2\pi)^4} e^{-iqx} {q^\mu\gamma_\mu\over q^2-s}
  ={x_\mu\gamma^\mu\over 4\pi^2 x^2} sK_2(\sqrt{s}x)
\end{equation}
for space-like $x$.
In practice we always normalize $R(x)$ by the corresponding
free correlator with massless quarks, $R_0(x)$,
\begin{equation}
R^{\rm meson}_0(x)={1\over x}\int_0^\infty ds\,
  f_0^{\rm meson}(s)\sqrt{s} K_1(\sqrt{s} x)\sim {1\over x^6} \, ,
\end{equation}
and
\begin{equation}
R^{\rm baryon}_0(x)=\int_0^\infty ds\, f_0^{\rm baryon}(s) s
K_2(\sqrt{s}x) \sim {1\over x^8} \, .
\end{equation}
An analogous normalization is also done for the lattice data,
as described more specifically in the next section.

In Fig.~\protect\ref{sketch}(b) we sketch the corresponding two-point function
in coordinate space, normalized by the free correlators with
massless quarks. To distinguish the physics in difference regions,
we plot the resonance and continuum contributions separately.
Due to asymptotic freedom, the continuum piece, denoted by the dashed
line, starts at 1 and then gradually decays to zero with
its width characterized by $\sim 1/\sqrt{s_0}$. In contrast,
the resonance starts as a power of $x$ at small $x$ and reaches
its maximum at around $\sim 1/M$ with a height proportional to
$\sim\lambda^2$. Although the full correlation function is the
sum of these two contributions, it is important to note that both in
this example and in many of the physical calculations it is possible to
distinguish the resonance and continuum contributions to a large extent
and thereby understand the physical origin of the numerical results.

  For later convenience we list all the spectral functions and
their corresponding parameterized correlators.

In the vector channel
\begin{equation}
f^V(s)=3\lambda^2_\rho\delta(s-M^2_\rho)+{3s\over 4\pi^2}\theta(s-s_0)\, ,
\end{equation}
\begin{equation}
R^V(x)/R_0^V(x) =
{\pi^2\over 8}\left( {\lambda_\rho\over M_\rho^2}\right)^2
(M_\rho x)^5 K_1(M_\rho x)
+{1\over 16} \int_{\sqrt{s_0}x}^\infty d\alpha\, \alpha^4 K_1(\alpha)\, .
\end{equation}
The factor of 3 in front of $\lambda^2_\rho$ is due to contraction
of indices ${\mu,\nu}$ from $q^2 g_{\mu\nu}-q_\mu q_\nu$.

In the axial vector channel
\begin{equation}
f^A(s)=3\lambda^2_{A_1}\delta(s-M^2_{A_1})-f_\pi^2 M_\pi^2\delta(s-M_\pi^2)
+{3s\over 4\pi^2}\theta(s-s_0)\, ,
\end{equation}
\begin{eqnarray}
R^A(x)/R_0^A(x) =& &
{\pi^2\over 8}\left( {\lambda_{A_1}\over M_{A_1}^2}\right)^2
(M_{A_1}x)^5 K_1(M_{A_1}x)
-{\pi^2\over 24}\left({f_\pi\over M_\pi}\right)^2
(M_\pi x)^5 K_1(M_\pi x) \nonumber \\
&+&{1\over 16} \int_{\sqrt{s_0}x}^\infty d\alpha\, \alpha^4 K_1(\alpha)\, .
\end{eqnarray}
In this channel there are two invariants; $q^2 g_{\mu\nu}-q_\mu q_\nu$
and $q_\mu q_\nu$ which have independent spectral functions; the first
includes the $A_1$ pole and the second includes the pion pole.
The relative minus sign between the $A_1$ and $\pi$ pole
terms, arising from kinematics, explains why the correlator in this
channel becomes negative at very large distances.
The continuum contribution from the two invariants have the same
asymptotic form and are expressed as a single term.

In the pseudoscalar channel
\begin{equation}
f^P(s)=\lambda^2_\pi\delta(s-M^2_\pi)+{3s\over 8\pi^2}\theta(s-s_0)\, ,
\end{equation}
\begin{equation}
R^P(x)/R_0^P(x) =
{\pi^2\over 12}\left( {\lambda_\pi\over M_\pi^2}\right)^2
(M_\pi x)^5 K_1(M_\pi x)
+{1\over 16} \int_{\sqrt{s_0}x}^\infty d\alpha\, \alpha^4 K_1(\alpha)\, .
\end{equation}

In the scalar (iso-vector) channel
\begin{equation}
f^S(s)={3s\over 8\pi^2}\theta(s-s_0)\, ,
\end{equation}
\begin{equation}
R^S(x)/R_0^S(x) =
{1\over 16} \int_{\sqrt{s_0}x}^\infty d\alpha\, \alpha^4 K_1(\alpha)\, .
\end{equation}
Since there is no known resonance in this channel, only the continuum
part is included in the spectral function.

In the nucleon channel
\begin{equation}
f^N(s)=q_\mu\gamma^\mu \left[\lambda^2_N\delta(s-M^2_N)
+{s^2\over 64\pi^4}\theta(s-s_0)\right]+\cdots\, ,
\end{equation}
where $\cdots$ represents terms with other Dirac structure and
\begin{equation}
R^N(x)/R^N_0(x) ={\pi^4\over 96}
\left({\lambda_N\over M_N^3}\right)^2 (M_N x)^8 K_2(M_N x)
  + {1\over 3072}
\int_{\sqrt{s_0}x}^\infty d\alpha\, \alpha^7 K_2(\alpha)\, .
\end{equation}

In the Delta channel
\begin{equation}
f^\Delta(s)=q_\mu\gamma^\mu \left[2\lambda^2_\Delta\delta(s-M^2_\Delta)
+{3s^2\over 256\pi^4}\theta(s-s_0)\right]+\cdots\, ,
\end{equation}
where again the $\cdots$ stands for terms with other Dirac structure and
\begin{equation}
R^\Delta(x)/R^\Delta_0(x) ={\pi^4\over 36}
\left({\lambda_\Delta\over M_\Delta^3}\right)^2
(M_\Delta x)^8 K_2(M_\Delta x)
  + {1\over 3072}
\int_{\sqrt{s_0}x}^\infty d\alpha\, \alpha^7 K_2(\alpha)\, .
\end{equation}
The factor of 2 in front of $\lambda^2_\Delta$ comes from $g_{\mu\nu}$
multiplying with the Rarita-Schwinger tensor
$$\left[g_{\mu\nu}-{1\over 3}\gamma_\mu\gamma_\nu
  +{1\over 3}(\gamma_\mu q_\nu-\gamma_\nu q_\mu)
  -{2\over 3}{q_\mu q_\nu \over M_\Delta^2 \,}\right]\, ,$$
with $q^2=M_\Delta^2$.

Because the local currents we have used are not conserved on the
lattice, we need, in principle, to know the relevant
finite renormalization factors before we can assign a definite
meaning to the fitted values of $\lambda$. Although in the mesonic channels
these renormalization factors were estimated in some cases in the
literature \cite{martinelli},
there  similar estimates in the baryonic
channels, as far as we know. To circumvent this problem, we
have fitted the pole term relative to the
continuum term and ignored the overall normalization. As long
as the currents are multiplicatively renormalized, which we
assume, this scheme should be a good way to avoid the explicit
renormalization factors. The primary limitation is that,
due to the relatively
poor knowledge of the continuum contribution $f_c(s)$, we introduce
a systematic error when we approximate $f_c(s)$ by
$f_0(s)$. However, we do not expect the systematic error to be
too large, since $f_c(s)$ does not have pronounced peaks near the
threshold. Ultimately, a consistency check should be made with
other fitting schemes.

\section{ LATTICE CALCULATION}

The lattice calculations were performed on a $16^3\times 24$ lattice in the
quenched approximation with Wilson fermions at an inverse coupling
$6/g^2=5.7$, corresponding to a physical lattice spacing defined by the proton
mass of approximately $a=0.17$~fm.  Lattice spacings defined by the string
tension or the rho mass would be 15\% higher or lower, respectively.  The
motivation for using such a large coupling, which is crude by current
standards, is the fact that the necessary propagators for point sources were
available from Soni {\it et. al.} \cite{soni},
while propagators for larger lattices
and correspondingly larger inverse coupling constant are normally calculated
from distributed sources to optimize the overlap with hadronic wave functions
and are thus unsuitable for the current application.  This inverse coupling
constant is large enough to give a semi-quantitative approximation and has
allowed us to make a thorough study of the finite lattice effects described
below.  Point propagators for 16 configurations were used for five values of
the hopping parameter, $\kappa = 0.154, 0.160, 0.164, 0.166,$ and $0.168$.
To think about these propagators in physical units,
it is convenient to
associate a quark mass $m_q \equiv \left( 1/2\kappa - 1/2\kappa_c\right)
a^{-1}$ with
each value of $\kappa$, where $\kappa_c = 0.1692$,
yielding the five values of $m_q$, 317, 199, 110, 67,
and
25~MeV, respectively.  Extrapolation to the value of $\kappa$ which
reproduces the
pion mass then corresponds to extrapolation to $m_q = 5.2$~MeV.  Because the
quark propagators had hard-wall boundary conditions in the time direction, all
correlation functions were calculated on the central time slice containing the
source.  Finite size effects in the spatial direction are discussed below.  In
order to calculate the ratio of interacting to free correlation functions,
corresponding free quark lattice propagators were calculated for $m_q a= 0.05$
on a $(48)^4$ lattice, which was sufficiently large to eliminate finite volume
effects for spatial separation less than 4~fm.  The effect of the small quark
mass $m_q$ was estimated in the continuum by evaluating the ratio of
non-interacting quark correlation functions
$\langle J\bar{J}\rangle\big|_m\big/\langle J\bar{J}\rangle\big|_0$,
with the result that the finite mass overestimated the
ratio at $x/a = 5$ by factors ranging from the smallest value of 0.3~\% in
the vector channel to the largest value of 5.5~\% in the axial channel.
Preliminary analysis of the same lattice calculation has been reported
in ref. \cite{letter}.

\subsection{ Lattice Anisotropy}
In order to obtain a physical approximation to the continuum correlation
functions, it is necessary to understand and correct for all relevant lattice
artifacts.  A particularly important lattice effect for the present
application is the anisotropy introduced into the rotationally invariant
continuum correlation functions by the Cartesian lattice.

The effect and a means for dealing with it are clearly displayed for the case
of non-interacting quarks on the lattice.  In the hopping-parameter expansion,
it is clear that points in Cartesian directions can be reached with fewer
steps than equidistant points in other directions. Although at very large
separations entropy effects should favor wiggly paths which approach the
continuum result in all directions, one clearly expects propagators and thus
correlation functions to be larger than the continuum result for points near
the Cartesian axes, and for this effect to become more pronounced at short
distances.  These expectations are clearly borne out in the case of the
vector meson correlation functions shown in Fig.~\ref{anisotropy_vector}.
The solid line in the
upper curve denotes the continuum result for non-interacting quarks.  All
discrete lattice results have been calculated on the central time slice (to
avoid complications from hard wall boundary conditions) and averaged over
equivalent permutations of the axes, so we will denote sites by $\left(
n_x,n_y,n_z\right)$ with $n_x\ge n_y\ge n_z$.  By the previous argument, the
Cartesian sites $(n,0,0)$ should lie further above the continuum, and as shown
by the diamonds in the figure, these points indeed do lie the highest.  The
diagonal directions ($n,n,n$) should suffer least from the lattice artifacts,
and we observe that these points, denoted by circles, do approximate the
continuum well throughout.  The sub-diagonal direction ($n,n,0$) denoted by
squares, is intermediate between the two extremes.

By asymptotic freedom, one would expect the QCD solution on the lattice to
approximate the free solution at short distances, and one observes from the
lower curve in the top portion of Fig.~\ref{anisotropy_vector}
that the qualitative structure of the
QCD solution, including its spatial anisotropies, is remarkably similar to the
free case plotted above it.  Note that for clarity, statistical error bars
have been omitted but are comparable to the symbol size.  We therefore have
every reason to believe that the diagonal sites should give a good
approximation to the continuum, and furthermore that much of the anisotropy
will in fact cancel out when we calculate the ratio of the QCD correlation
function to the free correlation function.

The ratio of the QCD to free correlation functions is plotted in the lower
section of Fig.~\ref{anisotropy_vector}.
Here, the error bars, which are still omitted for clarity,
are typically of the order of 2\% at $x/a=5$ and 15\% at  $x/a=10$.
In principle one would like to normalize the ratios at an
infinitesimally small separation. Although in
Fig.~\ref{anisotropy_vector}
we arbitrarily normalized the data at the closest point
(0,0,1), since we have argued that the diagonal points are most reliable,
we have normalized all our subsequent physical results at the
first non-zero diagonal separation $(1,1,1)$, corresponding to a physical
separation of $\sim 0.29$~fm.  Note for future reference that we are
considering the ratio of quantities which separately are varying over six
orders of magnitude.  As one would expect, at small $x$, the lattice
anisotropy is nearly identical in the interacting and free case so that the
ratio is nearly isotropic.  At intermediate distances $x/a\sim 5$, the
cancellation is not complete, but an order of magnitude effect in the QCD
correlation function is reduced to a 50\% effect in the ratio.  In this
respect the vector case we show in
Fig.~\ref{anisotropy_vector} is a worst case, and most other
channels have much greater cancellation.  At large $x$, the anisotropy
becomes negligible in both numerator and denominator.  In order to increase
the statistics, while maintaining a good approximation to the continuum, we
have adopted a prescription of including not only the diagonal sites ${\vec
d}\equiv (n,n,n)$ denoted by the circles, but also all sites ${\vec r} =
(n_x,n_y, n_z)$ such that $\hat r\cdot\hat d\ge 0.9$.  That is, we include all
sites falling within a cone surrounding the diagonal with opening angle
$\theta =\cos^{-1} (0.9)\approx26^\circ$.  The rationale for this prescription
is that propagators to these points sample the same general class of quark
paths as the diagonal points, and one observes that the fancy crosses in
Fig.~\ref{anisotropy_vector} denoting these sites indeed
define a smooth curve which includes the
circles representing the pure diagonal points.

The behavior shown in
Fig.~\ref{anisotropy_vector} is representative of all the meson and baryon
channels we have calculated.  In particular, data measured within the
$26^\circ$ cone surrounding the diagonal always lie on a universal curve
for all channels and the anisotropy associated with the Cartesian directions
is smaller in all other channels.
The analogous correlation functions for nucleon
currents are shown in Fig.~\ref{anisotropy_nucleon},
where all the symbols and plots are defined as in
Fig.~\ref{anisotropy_vector}.
The primary difference with respect to the meson case arises from the
fact that the correlation functions contain three quark propagators
and include the contraction $\gamma _\mu x^\mu$ so that the free case
falls like $x^{-8}$ instead of $x^{-6}$.  The anisotropies have the same
behavior as before, are extremely similar in the interacting and
non-interacting case, and cancel out even more completely than before in the
ratio.  The locus of points specified by circles and fancy crosses again
defines a smooth curve which we believe to be a good physical approximation to
the continuum.

\subsection{ Image Corrections}

Due to periodic boundary conditions for quarks at the spatial boundaries, the
large distance behavior of quark propagators is affected by the presence of
image sources in adjacent unit cells.  The correction for image s in
correlation functions of the form considered in this work is discussed in
detail in references \cite{image,grandy}.
The main point is that because
$\langle \Omega\left| J(x) \bar{J}(0)\right|\Omega\rangle$
involves a gauge-invariant closed loop of propagators,
any cross terms in which the contraction of one propagator involves
different fundamental and image currents than the other propagator necessarily
has the topology of a Wilson line encircling the entire lattice and thus is
negligible in the confining phase.  Hence, only diagonal terms involving
images occur, and the effect of all first images is to yield the desired
infinite volume correlation function summed over all first image sources.
Special symmetry points may be corrected trivially.  For example the center of
a face of the unit cell $\left( {N \over 2},0,0\right)$,
where $N$ is the linear
dimension of the unit cell (16 in our case), is equidistant from two sources
and the correlation function at this point is thus multiplied by 1/2.
Similarly, the center of an edge $\left( {N \over 2},
{N \over 2},0\right)$ is equidistant
from four sources yielding a factor $1/4$, and the corner $\left(
{N \over 2},{N \over 2},{N \over 2}\right)$ is equidistant from
eight sources yielding a factor of
$1/8$.  For all other points, one must numerically subtract the image
contributions, which is done iteratively by approximately correcting for
images using an appropriately defined parametric curve, least squares fitting
the parameters to the corrected data, and iterating to self-consistency.  In
practice, this procedure always yields a smooth universal curve at large
distances for the cases of interest here, and the lattice QCD data in
Fig.~\ref{anisotropy_vector}
and Fig.~\ref{anisotropy_nucleon} have been corrected in this way.

\subsection{ Extrapolation}

Because it is impractical to calculate quark propagators at a quark
mass light enough to produce a physical pion, it is necessary to
perform a sequence of calculations at a series of heavier quark masses
and extrapolate to the quark mass corresponding to the physical pion
mass.  Nearby lattice data are grouped in bins of $n \le x < n+1$
lattice spacings, and data within each bin are combined to a single
value by means of a statistically weighted average, with each
correlator datum $y_i$ given a weight $w_i = 1/\sigma_i^2$, where
$\sigma_i$ is the statistical uncertainty of the datum.
In the pseudoscalar channel, where the fits are
particularly sensitive to the abscissas, we determine these abscissas by
the statistically weighted average. In all other channels, the abscissas
of the binned data are approximated by the central values $n
+ 0.5$.  The binned data at each
separation, computed from the lightest four quark masses enumerated
above, are extrapolated using a least-squares quadratic fit.  In the
pseudoscalar channel we extrapolate the logarithm of the correlator,
and in the other channels the correlator itself is extrapolated.

 The extrapolation of the correlator in the pseudoscalar channel is
displayed in Fig.~\ref{pion_extrap}.
In this channel, the height of the resonance peak
diverges as $\lambda_\pi^2 / M_\pi^4$ and the position of the peak
also diverges as $1/M_\pi$. Despite these difficulties. we believe that
we have performed our calculations over a sufficient range of quark
masses to obtain a reasonable picture of the pion correlator at the
physical pion mass since the plot shows that our extrapolation is
indeed smooth.  The pseudoscalar channel is the worst case of the
extrapolation; the other channels contain no Goldstone bosons, and
consequently the extrapolations are well-behaved in the chiral limit.

We have shown the extrapolation as a function of the quark mass
in all the channels for two typical separations
in Fig.~\ref{kappa_extrap}. This figure shows
that the data at a given current separation
vary smoothly as a function of quark mass.  A quadratic fit to
all five data points gives a reasonable description of the mass
dependence over a range of 300 MeV, and the quadratic fit to
the lowest four masses used for the actual extrapolation
is seen to provide an excellent fit to all the data.
Thus we have confidence in
the extrapolation of the correlators from heavier quark masses to the
quark mass corresponding to the physical pion mass.

\subsection{Fitting with Phenomenological Spectral Function}

The hadronic current correlators for the six channels shown in Table I,
which have been computed for all geometrically
independent separations near the body diagonal and combined into bins
of one lattice unit of separation, are fitted using the dispersion
forms based on the phenomenological spectral functions described in
Section II.  Where possible, we have minimized $\chi^2$ with respect
to all four parameters: the arbitrary normalization, the
resonance mass $M$, the coupling $\lambda$ of the current to the resonance,
and the continuum threshold $s_0$.
As described below, in some channels the lattice
data do not allow us to freely fit all four parameters so we fix some
of the parameters to values independently known from previous lattice
calculations or phenomenology.  In the axial channel there are two
additional parameters associated with the pion resonance term, $f_\pi$
and $M_\pi$, and the fit is performed by allowing three of the six
available parameters to vary in the minimization of $\chi^2$.
As noted previously, we set the lattice
spacing to $a = 0.17\,{\rm fm}\,$ using the well-established value for
the proton mass \cite{ape,grandy}, with which our fitted proton mass is
consistent.

Using this fitting procedure, we extract the physical parameters of the
spectral functions. The
resonance masses, $M$, which are the energies of the lowest eigenstates
the quantum numbers of the current,
will be shown to be consistent with the masses measured accurately
by the asymptotic decay of two-point functions at large time
separations \cite{ape,grandy}.  From the ratio of the resonance to
continuum terms, we directly determine the physical coupling constants
which can be compared with experimental data and the
values used in existing models and sum rule
calculations. As discussed previously,
the threshold $s_0$ does not correspond precisely to the first excited
state in the data tables since it parameterizes the integrated strength
in the low energy region of the continuum. In the graphs presented below,
the correlators are
normalized to one at the separation $\sqrt{3}a = 0.29 {\rm fm}$.

\subsection{Error Analysis}

The lattice correlator measurements at different current separations at a
given quark mass are highly correlated, particularly at large
separations, and we compute the errors in the binned data directly
using the single elimination jackknife method.
In the quadratic extrapolation to
the quark mass $m_q=5.2 \,{\rm MeV}\,$, the uncertainty in the
extrapolated values is taken to be the range of the values for which
$\chi^2$ changes by less than one. Correlations between data for
different quark masses are not included in the calculation of
the standard deviations of the extrapolated data so it is possible
that these statistical uncertainties are underestimated.
Unfortunately, even with binned data, the correlation matrix is
so poorly conditioned that the
fitted parameters for the phenomenological spectral function
have to be
determined neglecting the correlations between the binned data
points.  The valley of $\chi^2$ in parameter space is in general rather
complicated, and for several channels and quark masses two nearby
minima of $\chi^2$ occur in parameter space.
Hence, we search directly for the range in each
fitted parameter for which a fit can be found that increases $\chi^2$
by less than one from the minimum value.  This search is accomplished
by repeatedly fixing one of the parameters in which $\chi^2$
is minimized and finding the minimum $\chi^2$ in the space of the
remaining parameters.  In general the
statistical range of each parameter is not symmetric about its best
fitted value, and we present optimal values and asymmetric ranges for
the fitted physical parameters.

\section{ Results}

This section presents the  results of our lattice calculations, their
parameterization in terms of the resonance and continuum spectral functions
described in Section II, and comparisons with phenomenological analyses
of experimental data and theoretical calculations based on QCD sum rules
and instanton models. The principal results for correlation functions
are presented in Figures 6-8 and the fitted parameters are tabulated
in Table II. We will present and discuss the results channel by channel.

\subsection{Vector Channel ($\rho$)}

The results for the vector channel are shown in the upper portion of
Figure 6, and provide a good example of a successful 4-parameter fit
to the lattice results.  As is evident from the figure, the continuum
term, denoted by the long dashes, and the resonance term, denoted by the
dotted curve, are separately well-determined and the sum fits the lattice
data, denoted by solid circles, quite well.

The $\rho$ mass extracted from this fit agrees well with the
mass measured from the exponential decay of the two-point function in
Euclidean time \cite{ape}. This agreement between the resonance mass
determined by the intermediate-range behavior of the correlation  function
and the asymptotic decay, which occurs systematically in all the channels we
have investigated and is a significant test of the consistency of the
parameterization of the spectral density, will be discussed subsequently
in connection with Figure 9. As is well known, however, at $\beta = 5.7$ the
$\rho$ mass on the lattice is lighter than the experimental mass.
As shown in Table II, both the values of the coupling constant $\lambda$ and
the threshold $s_0$ are close to the phenomenological values.

The lattice result for the vector correlation function is reasonably close
to the phenomenological result obtained by Shuryak \cite{shuryak}
from a dispersion
analysis of $e^+ e^- \rightarrow  $even number of $ \pi$'s. The fact that the
phenomenological
result lies below the lattice result follows from the fact that resonance
peak scales as $\lambda^2 / M_{\rho}^4$ and the lattice mass lies below
experiment while the coupling constant agrees with the phenomenological value.
The result of the instanton model is qualitatively similar, although lower
than phenomenology.

The most salient physics result in this channel is the fact that although
the free correlator falls by four orders of magnitude, the ratio of the
interacting to non-interacting correlators remains close to one. Although
the ratio must approach unity very close to $x=0$ by asymptotic freedom
and there is no leading order 't Hooft instanton induced interaction
in this channel, the ratio remains close to unity for much larger distances
than any simple arguments suggest.  This feature, which has been called
superduality, arises in this work as a ``conspiracy'' of the parameters
of the resonance and continuum terms of the spectral function.

\subsection{Pseudoscalar Channel ($\pi$)}

The pseudoscalar channel exhibits the most dramatic dependence on
the quark mass, reflecting the special role of the pion as a Goldstone boson.
In this case, the extrapolation is slightly sensitive to the fact we
extrapolated the log of the correlator, and logarithmic extrapolation in
x as well would give a slightly higher result.

The successful four-parameter fit shown in the lower panel of Fig. 6
provides strong support for our method of determining the resonance and
continuum terms. Note that because of the light pion mass,
the peak of the resonance occurs far outside
of the range in which the data is fit. Nevertheless, the extracted
mass and coupling constant agree well with the empirical results.
Because of the overlap of the resonance and continuum regions, the
threshold $\sqrt{s_0}$ is not fully determined. Rather, we only obtain a
bound of $1.0$ GeV by the criterion that $\chi^2$ increase by at most 1.
The fact that this is somewhat below the first excited state is
consistent with the fact that it must represent the integrated
strength included in the peaks of low-lying resonances.

As already noted in connection with Fig. 4, the lattice result is
close to the phenomenological result from the dispersion analysis
of ref \cite{shuryak}. It is also quite close to the results of the
random instanton vacuum model \cite {shuryak93} and
consequently the parameters in Table II agree well. Physically,
this is the most attractive channel and the leading order
't Hooft interaction is attractive in it.

\subsection { Scalar Channel}

The results in the scalar channel are shown in the upper panel of
Fig. 7. By the general dispersion analysis, the correlator should
be non-negative in this channel, and one observes that the lattice
results fall rapidly to zero at roughly 1 fm, albeit with large
errors at large distances. There is no evidence for resonances,
either experimentally or in the lattice calculation, and the
results are fit adequately by the continuum term with a single
threshold parameter.

Physically, since the 't Hooft interaction produces a repulsive
interaction from instantons to leading order, one expects a rapid
fall-off. However, the lattice results fall off much more slowly
than the random instanton vacuum model as shown in Fig 7.  The
overall behavior in the scalar channel is consistent with that
of two non-interacting ``very massive'' quarks.

\subsection { Axial Vector Channel ($A_1$)}

The axial vector channel is unique in that it is the only channel
for which the general dispersion analysis does not restrict the
spectral function to be positive. As shown in Eqs (2.13-2.14),
the axial and pseudoscalar contributions enter with opposite sign,
so we expect resonance terms of opposite sign in addition to
the continuum contribution.
As expected, one indeed observes that the lattice data in the
lower panel of Fig. 7 go negative at large distances. Although
one may wonder about the statistical significance given the large
error bars from extrapolation, the negativity is unambiguous in the
lattice measurements at each unextrapolated quark mass (see
for example the axial channel data at x=7.5 in Fig. 5.)
Because the data are insufficient to determine six parameters,
we fixed the masses of the $\pi$ and $A_1$, specified the
coupling of the $A_1$ to be the same as the $\rho$, and fit the
norm, threshold, and pion coupling.

In this channel, the extrapolated lattice result agrees quite well
with the phenomenological result derived partly from experimental
$\tau$--decay data \cite{shuryak} and with the results of the
random instanton vacuum model \cite{shuryak93}.

\subsection { Nucleon Channel }

The results for the nucleon channel are shown in the upper panel of
Fig. 8. In this case the lattice data have relatively small error
bars, and we obtain a good fit yielding  the correct nucleon mass.
The fit is relatively insensitive to the threshold $s_0$, and the
fit shown in the figure is
performed with $\sqrt{s_0}= 1.2$ GeV. The upper bound of
1.4 GeV in Table II is determined by the highest value
for which $\chi^2$ is increased by less than 1. In connection with
Fig. 5, one should note that, with the exception of the pseudoscalar
channel, the nucleon correlator has the strongest quark mass dependence
of any channel.

One observes that the lattice results are quite consistent with the
sum rule result of ref. \cite{ioffe}, shown by the dot-dot-dash curve.
In addition, although there are substantial statistical errors at large
distance, the random instanton vacuum model is also close to the
lattice results \cite{shuryak93}.

\subsection {Delta Channel}

The results in the delta channel are shown in the lower panel of Fig. 8.
In this case, we obtain a good four-parameter fit, with well-determined
continuum and resonance contributions shown in the figure. The mass
is slightly higher than the APE result, and will be discussed in connection
with Fig. 9 below. The coupling constant is consistent with the value
determined from the sum rule.

The results of the random instanton vacuum model are qualitatively
similar, but do not display as pronounced a resonance term and
also fall off more rapidly at large distance.  In plotting the
sum rule result in this channel, denoted by the dot-dot-dash line,
we have used the mass determined from the sum rule analysis itself,
rather than the experimental mass, to make the theory internally
consistent. This has the effect of reducing the height of the
resonance peak somewhat from ref. \cite{shuryak}.
Note that in the trace we have
calculated for our correlator given in Table II, there are spin
$1\over 2$ contaminants, so one should not expect complete agreement with
the sum rule result. To the extent that excited states are heavy,
the effect of these contaminants should not be too large.

\subsection {Mass Dependence of Parameters}

Having discussed the parameters characterizing the spectral function
for the fits to  extrapolated lattice data in each channel, it is
useful to observe the dependence of these parameters on the quark
mass. Hence, we have fit the correlators calculated for each quark
mass, and summarized the results in Figs. 9 and 10.

Figure 9 shows the resonance mass as a function of quark mass
for the vector, pseudoscalar, delta and nucleon channels. For
comparison, the mass dependence determined by the APE collaboration
\cite{ape} from extremely accurate measurements of the exponential
decay of two point functions at large time separations is also shown
by the solid curves. The detailed agreement between the fits to
the resonance masses and the APE results is striking and provides a strong
confirmation of the consistency and effectiveness of the parameterization
and fitting procedure. The agreement of the pion mass is particularly
significant, since the peak of the resonance is not even contained in
the region of the fit.  One statistical fluke worth noting is the
fact that the delta mass is slightly high at $m_q = 25$ MeV, which
also carries over into a high value for the extrapolated data.

Figure 10 shows comparable mass dependence for the coupling constants.
Although there is no smooth reference curve in this case, one observes
that the coupling constants vary quite smoothly with the quark mass,
so that there is nothing pathological happening in the chiral limit.
For reference, the experimental or model results from Table II are
denoted by arrows at the physical pion mass.

\section {conclusions}

In summary, we believe these results demonstrate the feasibility and
utility of lattice calculations of these vacuum correlation functions
and of the phenomenological analysis of the results.

Even for the relatively large lattice spacing in this work, we have
understood and controlled the lattice artifacts associated with the
finite volume and anisotropy of the lattice. For those channels in
which empirical results are available from dispersion analysis of
experimental data, we have shown that our results are in
semi-quantitative agreement with experiment.

The phenomenological analysis of the lattice results in terms of
a four-parameter characterization of the spectral function has
been shown to be successful in parameterizing the data and in
understanding its physical content. The analysis has been shown
to be reliable in the sense that the fit parameters are systematically
consistent with other lattice measurements of masses and with phenomenology.

These results strongly motivate more definitive calculations on larger
lattices with $6/g^2 = 6$. Having established the reliability of lattice
results in channels for which experimental data exist, we may then use the
lattice calculation as a tool to study correlation functions in channels
in which experimental measurement are not feasible or are unavailable.
In addition, it is instructive to perform companion calculations
with cooled configurations in which the contributions of instantons
alone can be compared with the full lattice results and instanton
based models. Work in these directions is in progress.

{\bf Acknowledgements}

 It is a pleasure to thank  Edward Shuryak for extensive discussions
on many aspects of this work and for making data available to us
prior to publication. We wish to acknowledge a helpful discussion
with Boris Ioffe concerning application of sum rules to the delta
channel. We also thank Amarjit Soni for making his point propagators
available to us , and the National Energy Supercomputer Center for
providing Cray-2 computer resources.
This work is supported in part by funds
provided by the U. S. Department of Energy (D.O.E.) under contracts
\#DE-AC02-76ER03069 and \#DE-FG06-88ER40427, and the National Science
Foundation under grant \#PHY~88-17296.

\begin{figure}
\caption{ Sketch of generic spectral function and its Fourier transform.
The light solid curves in the left panel (a) show a typical spectral
function as a function of the mass squared, $s$, comprising an isolated
resonance and a region of broad overlapping resonances merging with
the continuum.  The heavy lines show the parameterization,
Eq.~(\protect\ref{spectral}), where the isolated resonance is
represented by a delta-function
at $M^2$ with strength $\lambda^2$ and the continuum is parameterized
by $f_0(s)$ for the non-interacting system (in this case linear in $s$)
with a cutoff, $s_0$.  The right panel (b) shows the coordinate space
Fourier transform of the parameterized spectral function.  The resonance
delta-function produces the broad solid curve whose position is proportional
to $M^{-1}$ with height proportional to $\lambda^2$. The continuum term
produces the dashed curve which is normalized to one at the origin with
extent proportional to $s_0^{-{\frac{1}{2}}}$.}
\label{sketch}
\end{figure}

\begin{figure}
\caption{ Vector meson correlation functions.  The upper curve shows the
correlation function for non-interacting massless quarks calculated in the
continuum (solid curve) and lattice results for $ma=0.05$.  Diamonds denote the
Cartesian directions $(n,0,0)$, squares denote the directions ($n,n,0)$,
circles denote the diagonal directions $(n,n,n)$ and the crosses denote all
other points.  The lower curve in the upper panel shows the corresponding
lattice results for interacting quarks with $m_q =110$ MeV, shifted by an
arbitrary normalization factor.  There is no longer
an exact continuum curve, but all other symbols have the same meaning and
clearly show the similarity of lattice artifacts in the free and
interacting cases.  The lower panel shows the ratio of interacting to free
correlation functions normalized to 1 at the first point,
where the circles, squares and diamonds denote the same
lattice directions as above. The fancy crosses denote all other lattice points
lying within a cone of opening angle $26^\circ$ surrounding the diagonal
directions.  The smooth continuous curve defined by these fancy crosses and the
circles represents the physical ratio of correlation functions.}
\label{anisotropy_vector}
\end{figure}

\begin{figure}
\caption{ Nucleon correlation functions.  All quantities are defined as in
Fig.~2.}
\label{anisotropy_nucleon}
\end{figure}

\begin{figure}
\caption{ Quark mass dependence of pseudoscalar correlation function. Open
circles denote binned lattice data at five values of the quark mass.
Extrapolation to the quark mass 5.2 MeV, which corresponds to the physical
pion mass, yields the points denoted by solid dots with the associated
statistical errors.
The solid lines are three-parameter fits to the data, and the dot-dashed line
indicates the phenomenological result of
Ref.~\protect\cite{shuryak}.
}
\label{pion_extrap}
\end{figure}

\begin{figure}
\caption{Extrapolation of correlation functions in the quark mass for
nucleon (N), delta (D), scalar (S), axial (A), vector (V), and
pseudoscalar (P) channels.
Binned lattice data for $x$ in the intervals of 3-4 lattice units and
7-8 lattice units are shown as a function of the quark mass in MeV.
Where error bars are not visible, they are smaller than the dot size.
Quadratic least squares fits to the last 4 and 5 data points are
shown by solid and dashed lines respectively.}

\label{kappa_extrap}
\end{figure}

\begin{figure}
\caption{Vector $(V)$ and
Pseudoscalar $(P)$ correlation functions are shown in
the upper and lower panels respectively.  Extrapolated lattice data are
denoted by the solid points with error bars.  Fits to the lattice data using
the phenomenological form discussed in the text are given by the solid curves,
with the continuum and resonance components denoted by short dashed and dotted
curves respectively.  The empirical results determined by dispersion analysis
of experimental data in ref. \protect\cite{shuryak}
are shown by the long dashed curves.  The open
circles denote the results of the random instanton vacuum model of
ref. \protect\cite{shuryak93}. }
\label{result_vp}
\end{figure}

\begin{figure}
\caption{Scalar $(S)$ and Axial Vector $(A)$ correlation functions
are shown in the upper and lower panels respectively.  Extrapolated lattice
data and empirical results from
dispersion analysis of experimental data are given by solid dots
and long dashes as in Fig. 6.
The results of the random instanton vacuum model are denoted by
open circles which have been shifted slightly to the right where necessary
for clarity. The fit to the scalar lattice data includes
only a continuum term, denoted by the solid curve.  The fit to the axial
lattice data includes $A_1$ and $\pi$ resonance terms of opposite signs,
denoted by the dotted curves as well as the short dashed continuum curve,
yielding the total result given by the solid curve.}
\label{result_sa}
\end{figure}

\begin{figure}
\caption{Nucleon (N) and Delta $(D)$ correlation functions are
shown in the upper and lower panels respectively.  As in Fig.~6, extrapolated
lattice data are denoted by solid dots, the phenomenological fit is given by
the solid curve with continuum and resonance components given by dashed and
dotted curves respectively. The results of the random instanton model are
given by the open circles,
again shifted slightly where necessary for clarity.
The results from the QCD sum rule calculation of
ref. \protect\cite{ioffe} are indicated by the dot-dot-dashed lines.}
\label{result_nd}
\end{figure}

\begin{figure}
\caption{ Comparison of the masses extracted from the resonance term
in the spectral function, denoted by solid points with error bars, with masses
determined from the asymptotic decay of two-point functions by the APE
collaboration \protect\cite{ape},
shown by the solid curves.  The systematic agreement as a
function of quark mass $m_q$ in the vector $(V)$, pseudoscalar $(P)$, delta
$(D)$ and nucleon $(N)$ channels is a significant consistency check of the
present analysis.}
\label{mass_extrap}
\end{figure}

\begin{figure}
\caption{Dependence of the coupling constant $\lambda$ for the
resonance term on the quark mass, $m_q$~, in the vector $(V)$, pseudoscalar
$(P)$, delta $(D)$ and nucleon $(N)$ channels.  The five data points to the
right indicate lattice measurements which clearly extrapolate smoothly to the
chiral limit.  The left point denotes the results extrapolated
to $m_q=5.2$~MeV,
the quark mass corresponding to the physical pion.  Note that these results
compare well with the phenomenological results denoted by the arrows for the
vector and pseudoscalar channels and with the sum rule results denoted by the
arrows for the nucleon and delta channels.}
\label{lambda_extrap}
\end{figure}

\begin{table}
\caption{ Hadron Currents and Correlation Functions}
\begin{tabular}{llcc}
Channel & Current & Correlator $R(x)$ & $f_0(s)$ \\
\tableline
Vector & $J_\mu=\bar{u}\gamma_\mu d$ &
  $\langle\Omega|TJ_\mu(x)\bar{J}_\mu(0)|\Omega\rangle$ &
  $1/(4\pi^2)$ \\
\tableline
Axial  & $J_\mu^5=\bar{u}\gamma_\mu \gamma_5 d$ &
  $\langle\Omega|TJ_\mu^5(x)\bar{J}_\mu^5(0)|\Omega\rangle$ &
  $1/(4\pi^2)$ \\
\tableline
Pseudoscalar & $J^p=\bar{u}\gamma_5 d $ &
  $\langle\Omega|TJ^p(x)\bar{J}^p(0)|\Omega\rangle$ &
  $3s/(8\pi^2)$ \\
\tableline
Scalar & $J^s=\bar{u} d $ &
  $\langle\Omega|TJ^s(x)\bar{J}^s(0)|\Omega\rangle$ &
  $3s/(8\pi^2)$ \\
\tableline
Nucleon & $J^N=\epsilon_{abc}[u^aC\gamma_\mu u^b]\gamma_\mu\gamma_5 d^c$ &
  $\displaystyle{1\over 4}{\rm Tr}(
   \langle\Omega|TJ^N(x)\bar{J}^N(0)|\Omega\rangle x_\nu\gamma_\nu)$ &
  $s^2/(64\pi^4)$ \\
\tableline
Delta & $J^\Delta_\mu=\epsilon_{abc}[u^aC\gamma_\mu u^b] u^c$ &
  $\displaystyle{1\over 4}{\rm Tr}(
   \langle\Omega|TJ^\Delta_\mu(x)\bar{J}^\Delta_\mu(0)|\Omega\rangle
   x_\nu\gamma_\nu)$ &   $3s^2/(256\pi^4)$ \\
\end{tabular}
\label{table1}
\end{table}

\begin{table}
\caption{ Fitted Parameters}
\begin{tabular}{lllcc}
Channel & Source & M (GeV) & $\lambda$ & $\sqrt{s_0}$ (GeV)\\
\tableline
Vector & lattice & $0.72\pm 0.06$ & $(0.41\pm 0.02\,GeV)^2$ & $1.62\pm 0.23$\\
       & instanton\tablenotemark[1] &
         $0.95\pm 0.10$ & $(0.39\pm 0.02\, GeV)^2$ & $1.50\pm 0.10$ \\
       & phenomenology\tablenotemark[2] &
         0.78 & $(0.409\pm 0.005GeV)^2$ & $1.59\pm 0.02$ \\
\tableline
Pseudoscalar & lattice & $0.156\pm 0.01$ & $(0.44\pm 0.01\,GeV)^2$ & $<1.0$\\
       & instanton\tablenotemark[1] &
         $0.142\pm 0.014$ & $(0.51\pm 0.02\,GeV)^2$ & $1.36\pm 0.10$ \\
       & phenomenology\tablenotemark[2] &
         0.138 & $(0.480 GeV)^2$ & $1.30\pm 0.10$ \\
\tableline
Nucleon & lattice & $0.95\pm 0.05$ & $(0.293\pm 0.015\,GeV)^3$ & $<1.4$\\
       & Sum Rule\tablenotemark[3] &
         $1.02\pm 0.12$ & $(0.324\pm 0.016\,GeV)^3$ & $1.5$ \\
       & phenomenology\tablenotemark[2] &
         0.939 & ? & $1.44\pm 0.04$ \\
\tableline
Delta & lattice & $1.43\pm 0.08$ & $(0.326\pm 0.020\,GeV)^3$ & $3.21\pm 0.34$\\
       & Sum Rule\tablenotemark[3] &
         $1.37\pm 0.12$ & $(0.337\pm 0.014\,GeV)^3$ & $2.1$ \\
       & phenomenology\tablenotemark[2] &
         1.232 & ? & $1.96\pm 0.10$ \\
\end{tabular}
\tablenotetext[1]{Instanton Liquid Model \cite{shuryak93}.}
\tablenotetext[2]{Phenomenology estimated by Shuryak \cite{shuryak}
and from the particle data book \protect\cite{databook}.}
\tablenotetext[3]{QCD sum rule by Belyaev and Ioffe \cite{ioffe}.}
\label{table2}
\end{table}


\begin{references}

\bibitem[*]{byline} This work is supported in part by funds
provided by the U. S. Department of Energy (D.O.E.) under contracts
\#DE-AC02-76ER03069 and \#DE-FG06-88ER40427, and the National Science
Foundation under grant \#PHY~88-17296.

\bibitem{shuryak} E.~Shuryak, {\it Rev.~Mod.~Phys.\/} {\bf 65}, 1(1993).

\bibitem{martinelli} G.~Martinelli,  C.~T.~Sachrajda, and
A.~Vladikas, Nucl.~Phys.~{\bf B358}, 212(1991).

\bibitem{soni} A.~Soni, {\it National Energy Research Supercomputer Center
Buffer} {\bf 14}, 23 (1990); C. Bernard, T. Draper, G. Hockney, and A. Soni,
Phys Rev. {\bf D 38}, 3540 (1988).

\bibitem{letter}M.-C. Chu, J.~M.~Grandy, S.~Huang and J.~W.~Negele,
Phys.~Rev.~Lett.~{\bf 70}, 225(1993).

\bibitem{image}M.~Burkardt, J.~M.~Grandy and J.~W.~Negele,
MIT preprint CTP\#2108 (1993).

\bibitem{grandy}J.~M.~Grandy, {\it Investigation of Hadronic Structure by
Solving QCD on a Lattice}, Ph.~D. Thesis, Massachusetts Institute of
Technology (unpublished) (1992).

\bibitem{ape}Bacilieri, {\it et al. \/} (The APE Collaboration),
Nucl.~Phys.~{\bf B317}, 509(1989).

\bibitem{shuryak93}E.~V.~Shuryak and J.~J.~M. Verbaarschot, {\it Mesonic
Correlation Functions in the Random Instanton Vacuum},
Stony Brook Preprint SUNY-NTG-92-40 (Dec.~1992).

\bibitem{ioffe}B.~L.~Ioffe, {\it Nucl.~Phys.\/} {\bf B188}, 317 (1981);
V.~M.~Belyaev and B.~L.~Ioffe, {\it Sov.~Phys.~JETP} {\bf 83}, 976 (1982).

\bibitem{databook}{\it Review of Particle Properties}, Phys.~Rev.~{\bf D45},
Part 2(June 1992).






\end{references}
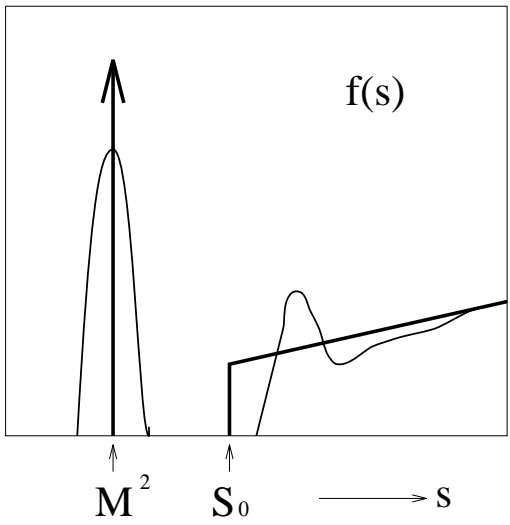
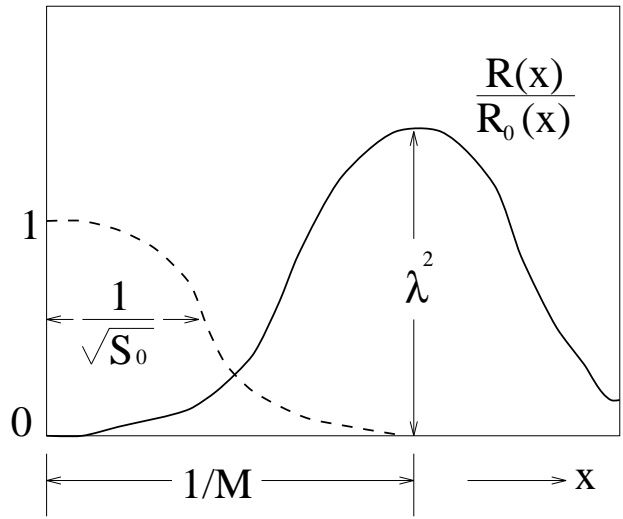
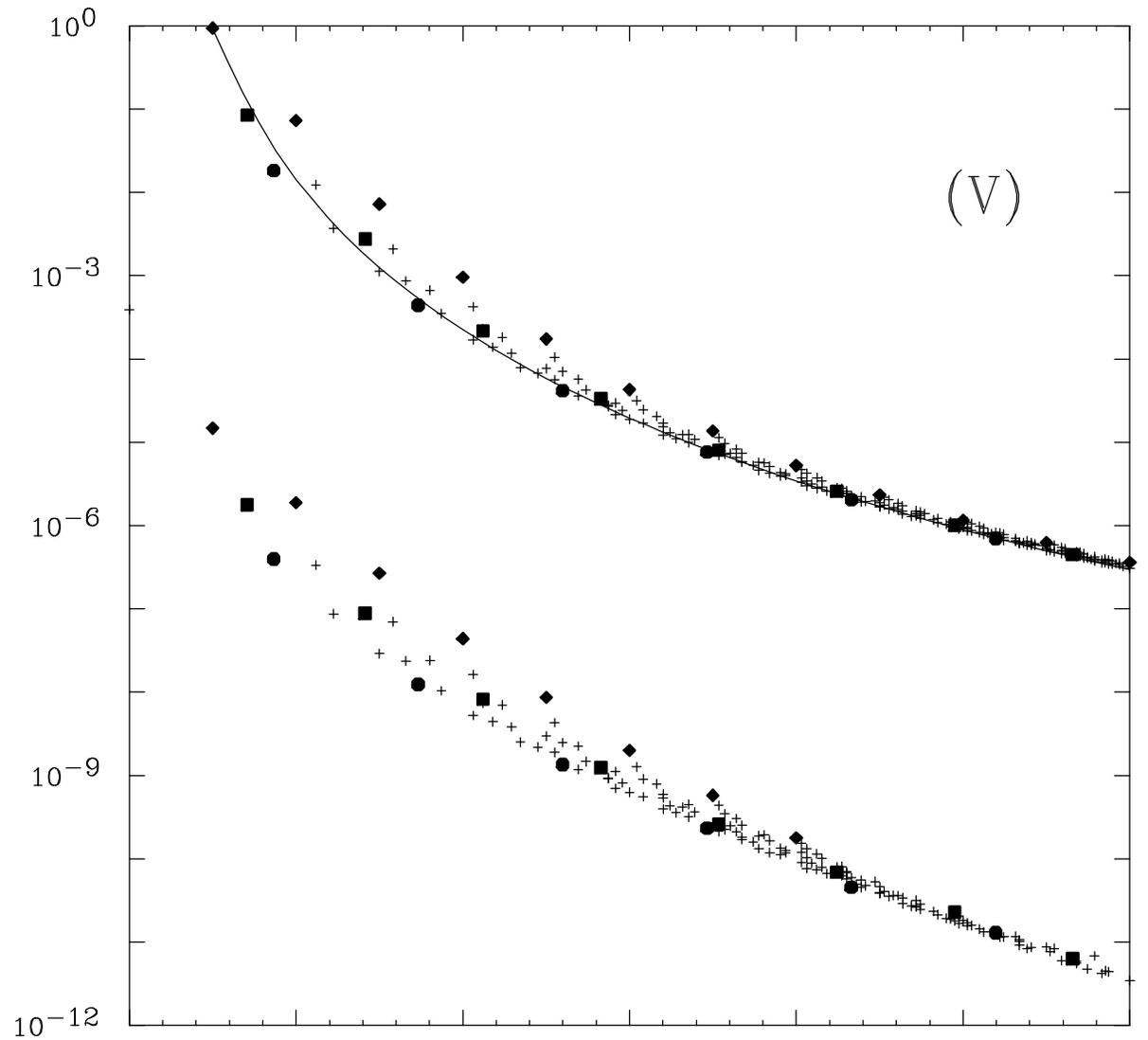
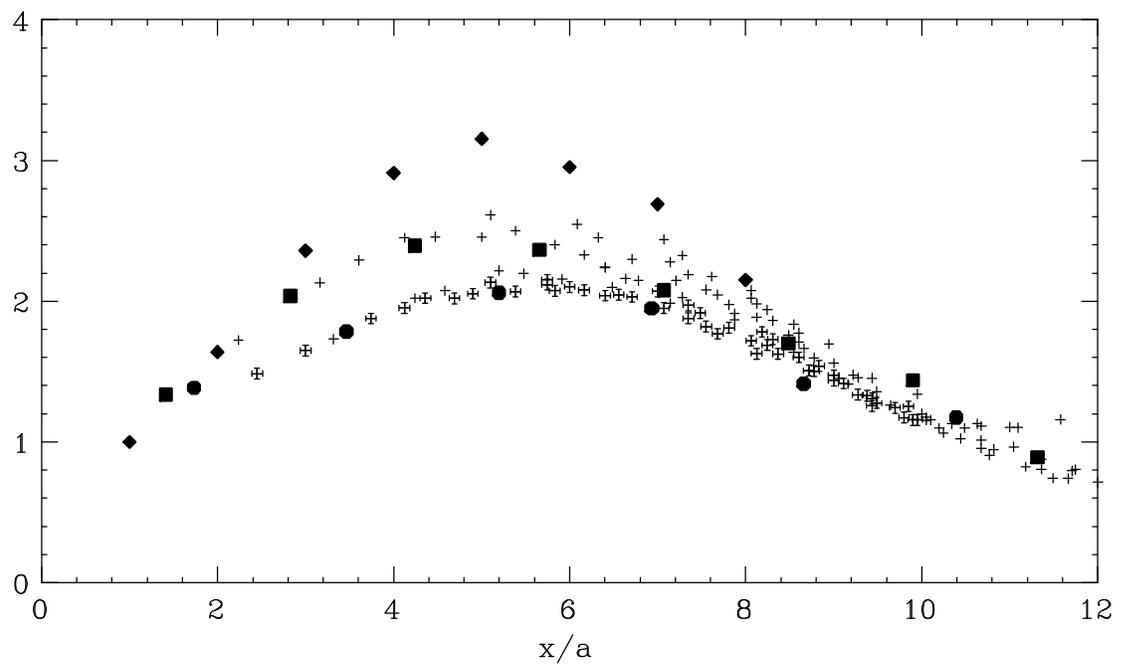
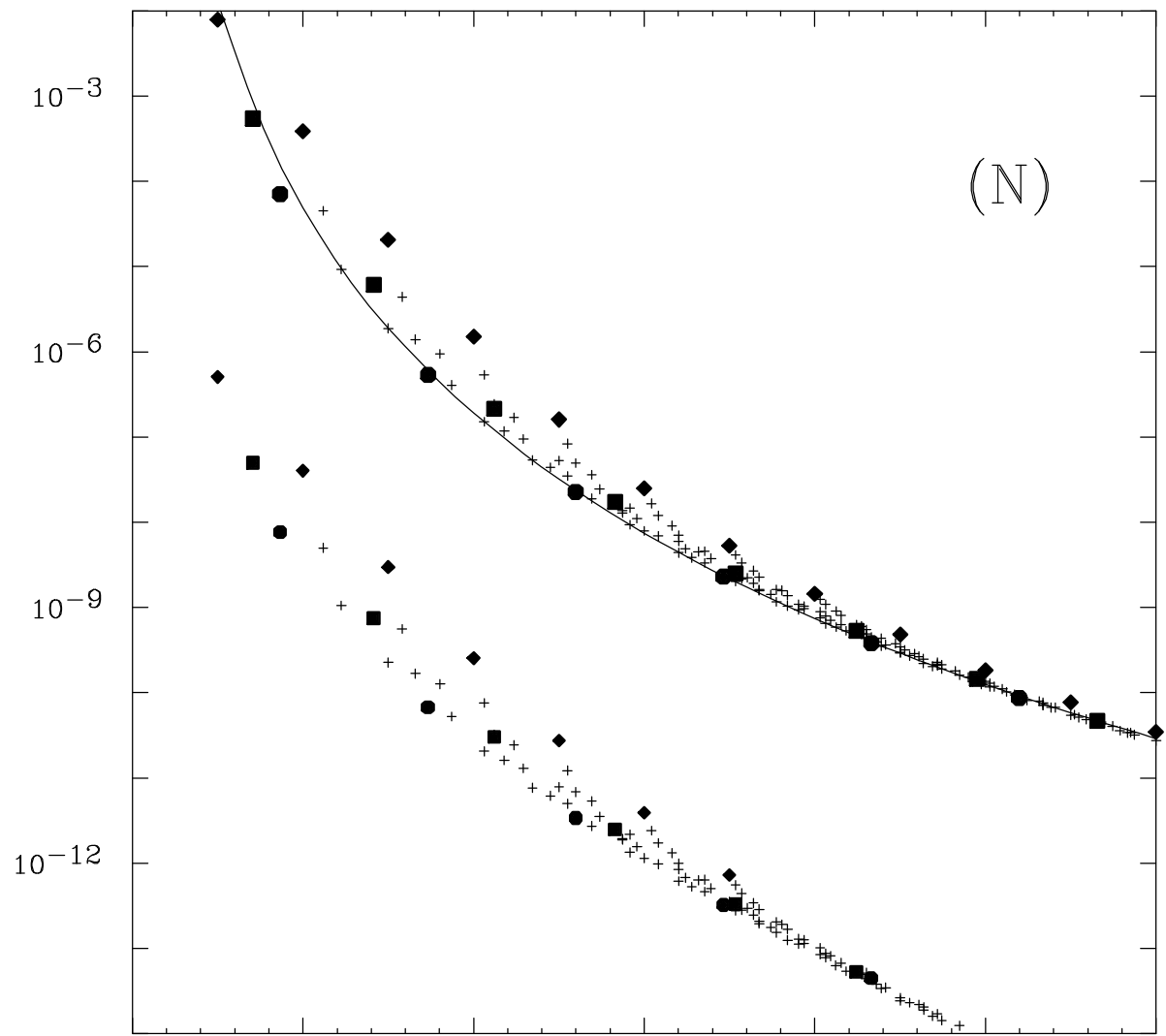
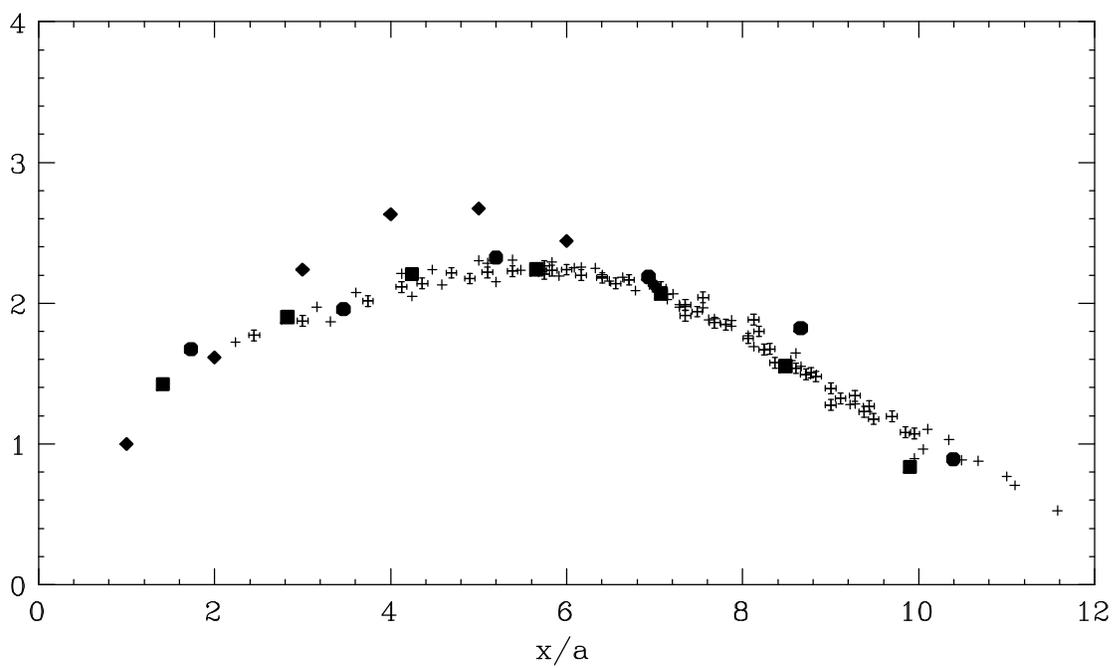
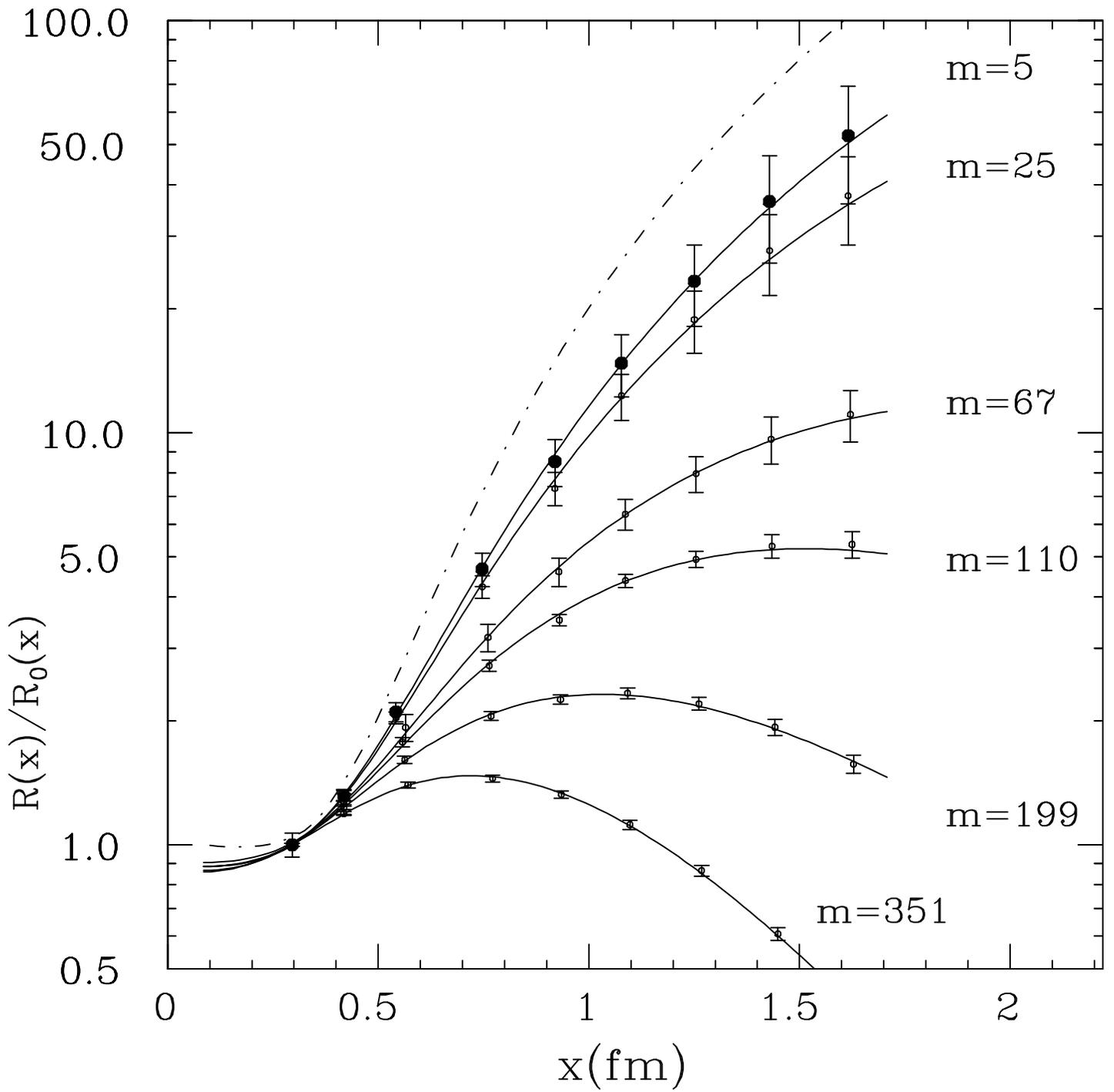
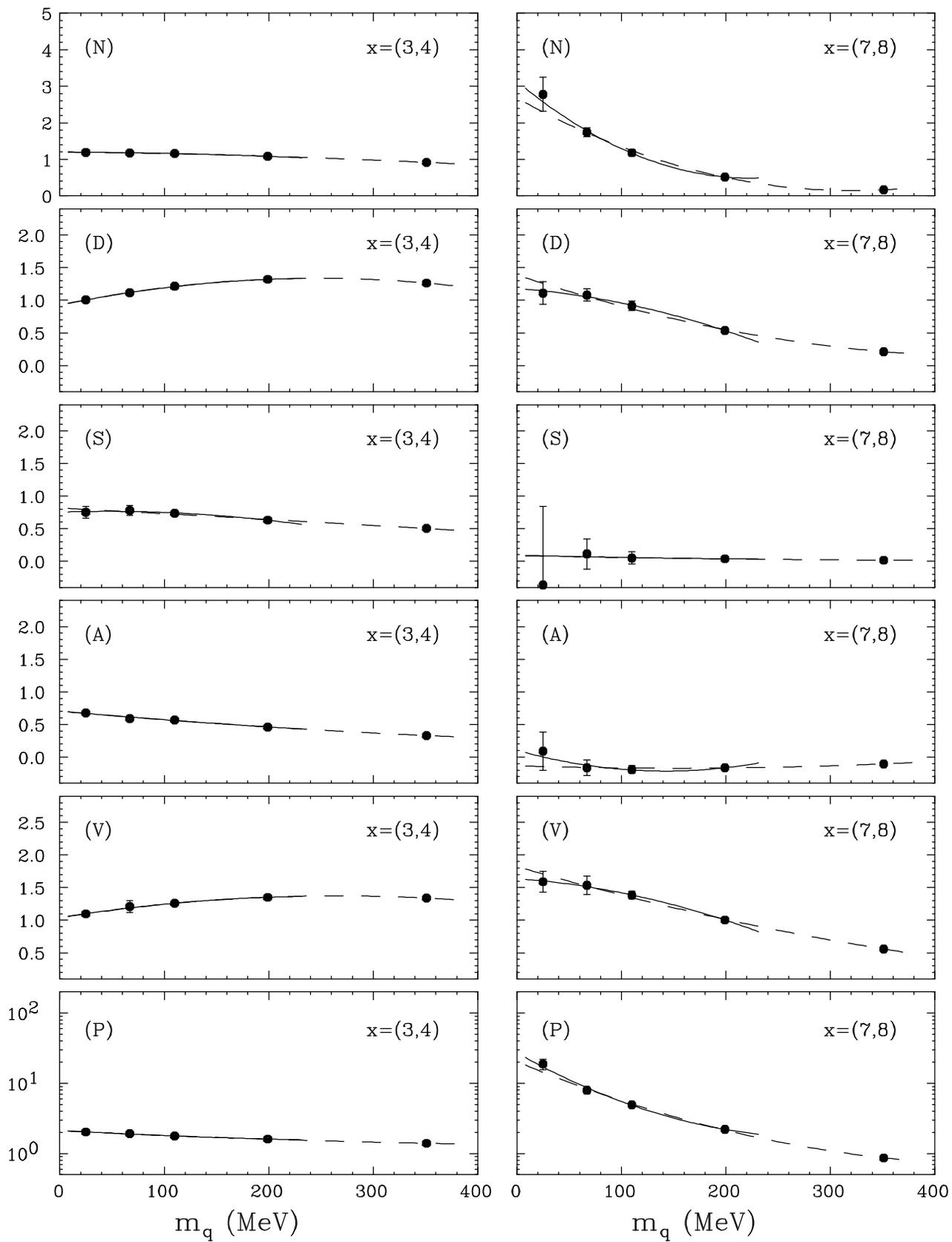
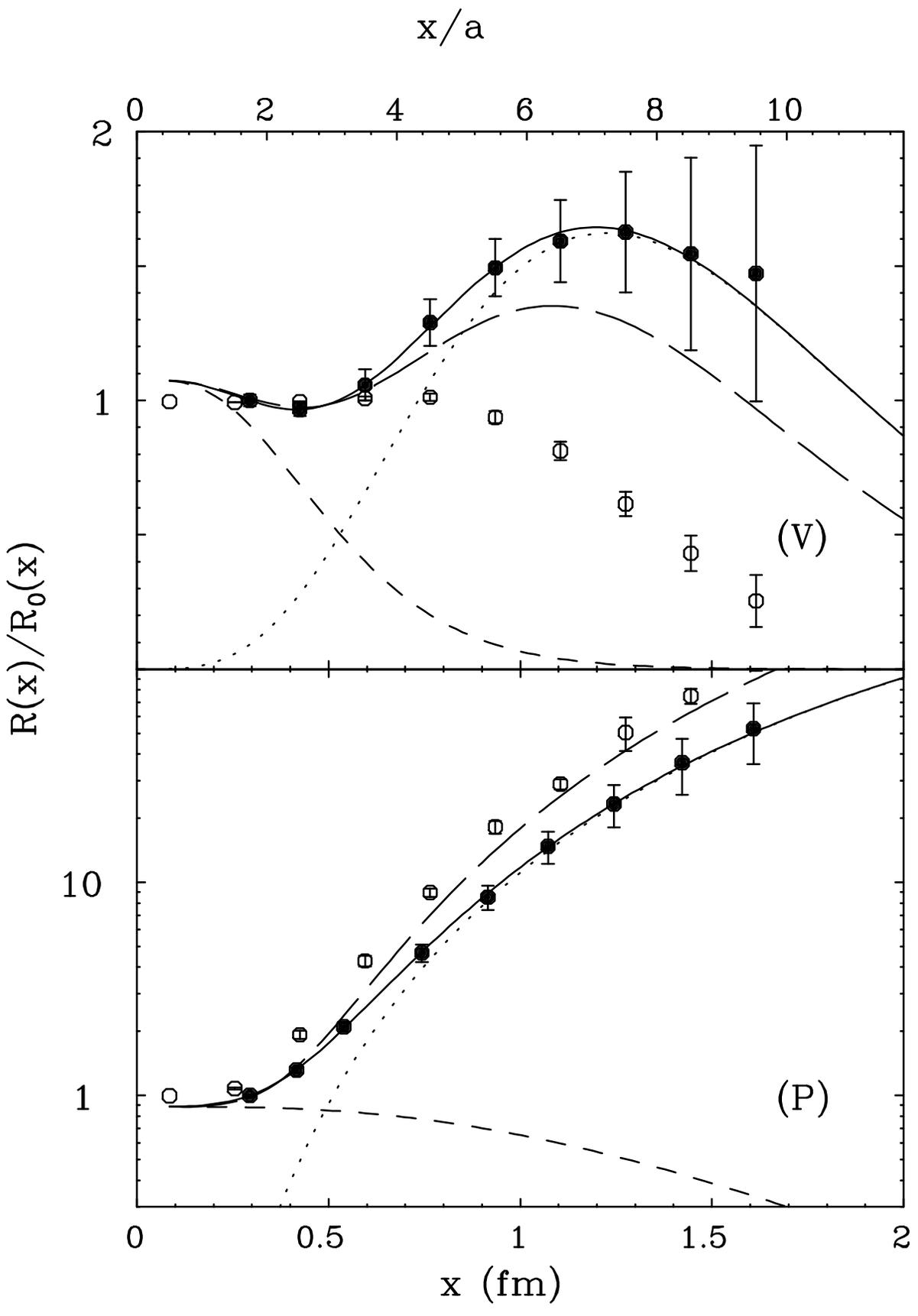
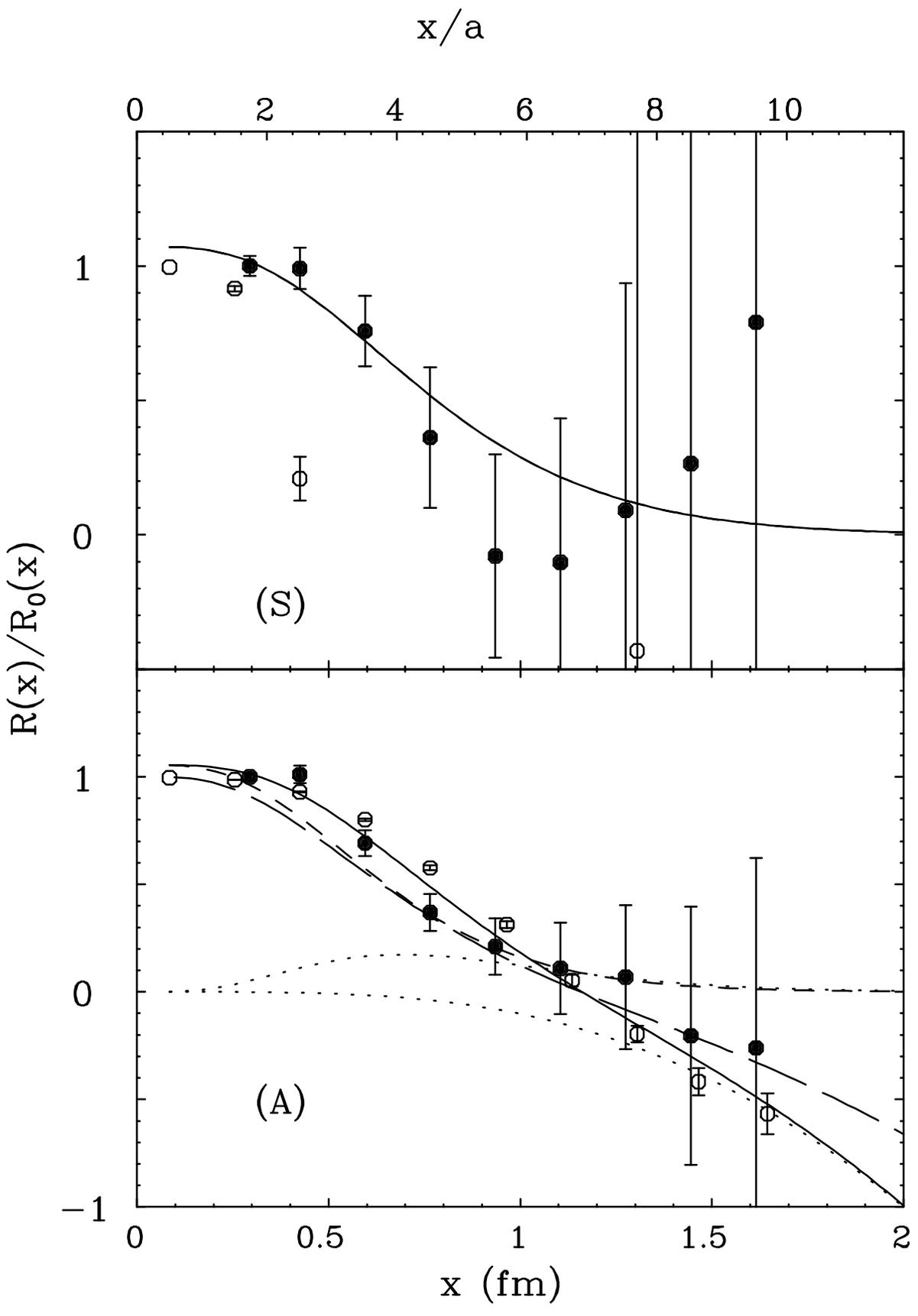
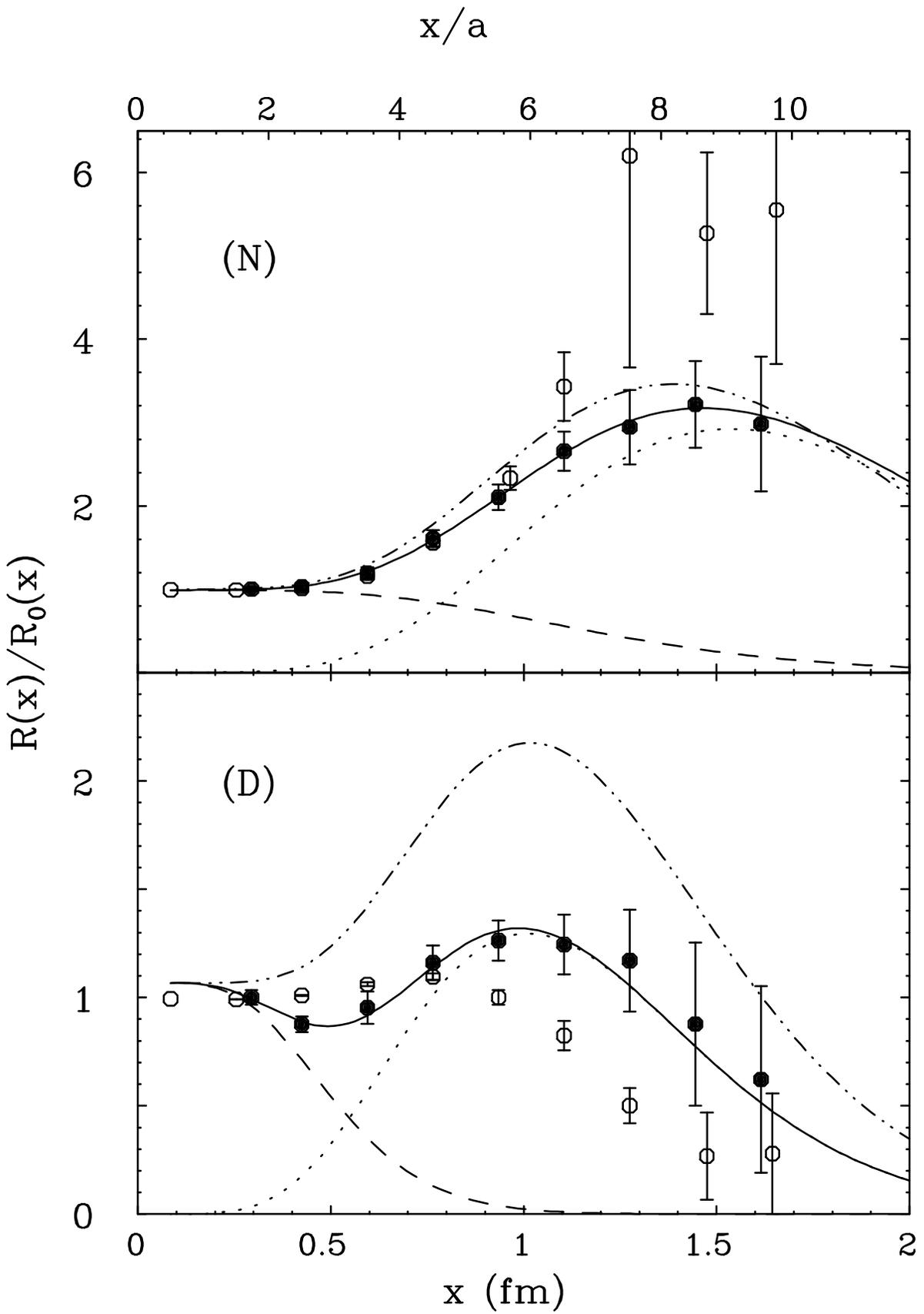
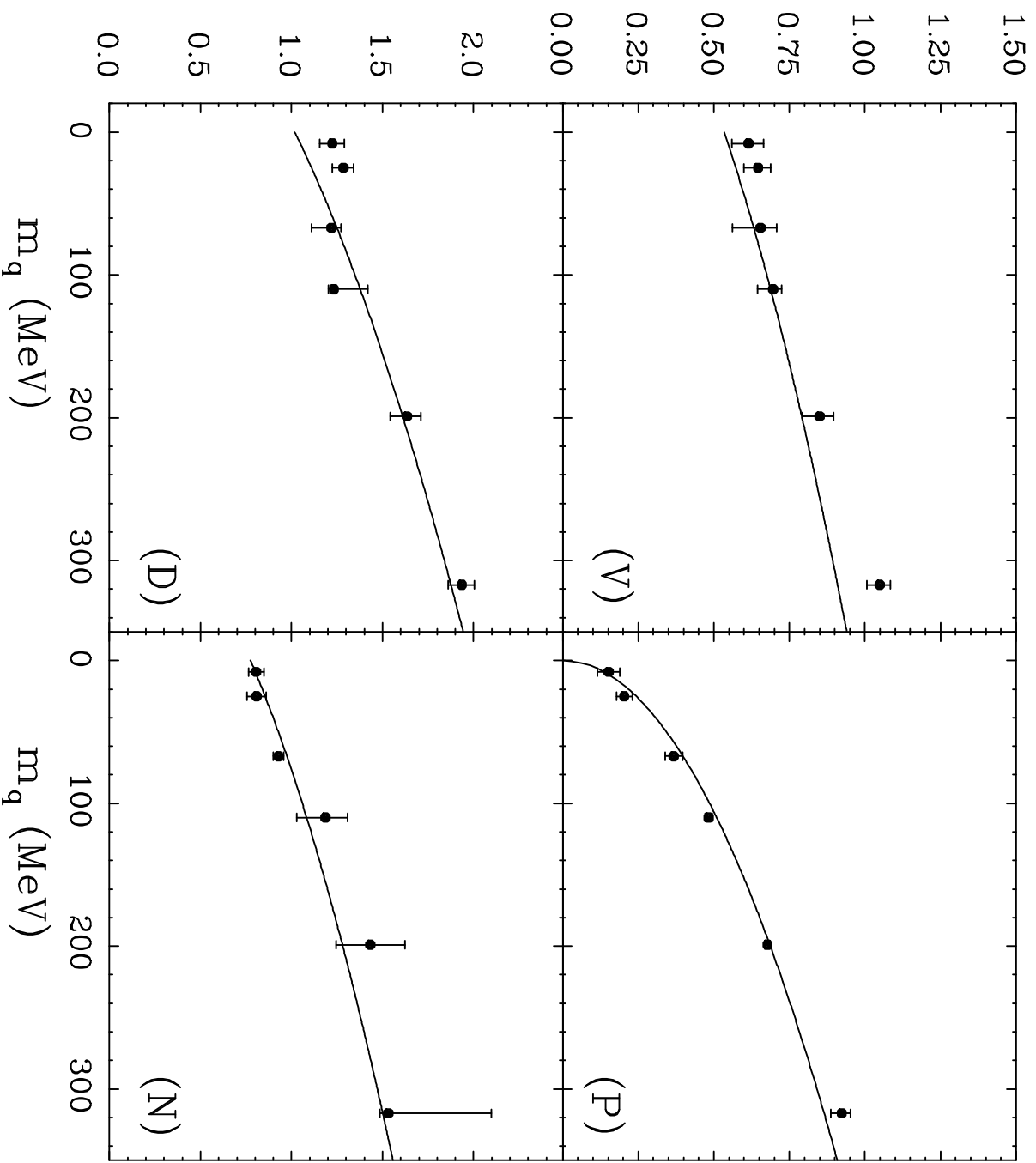
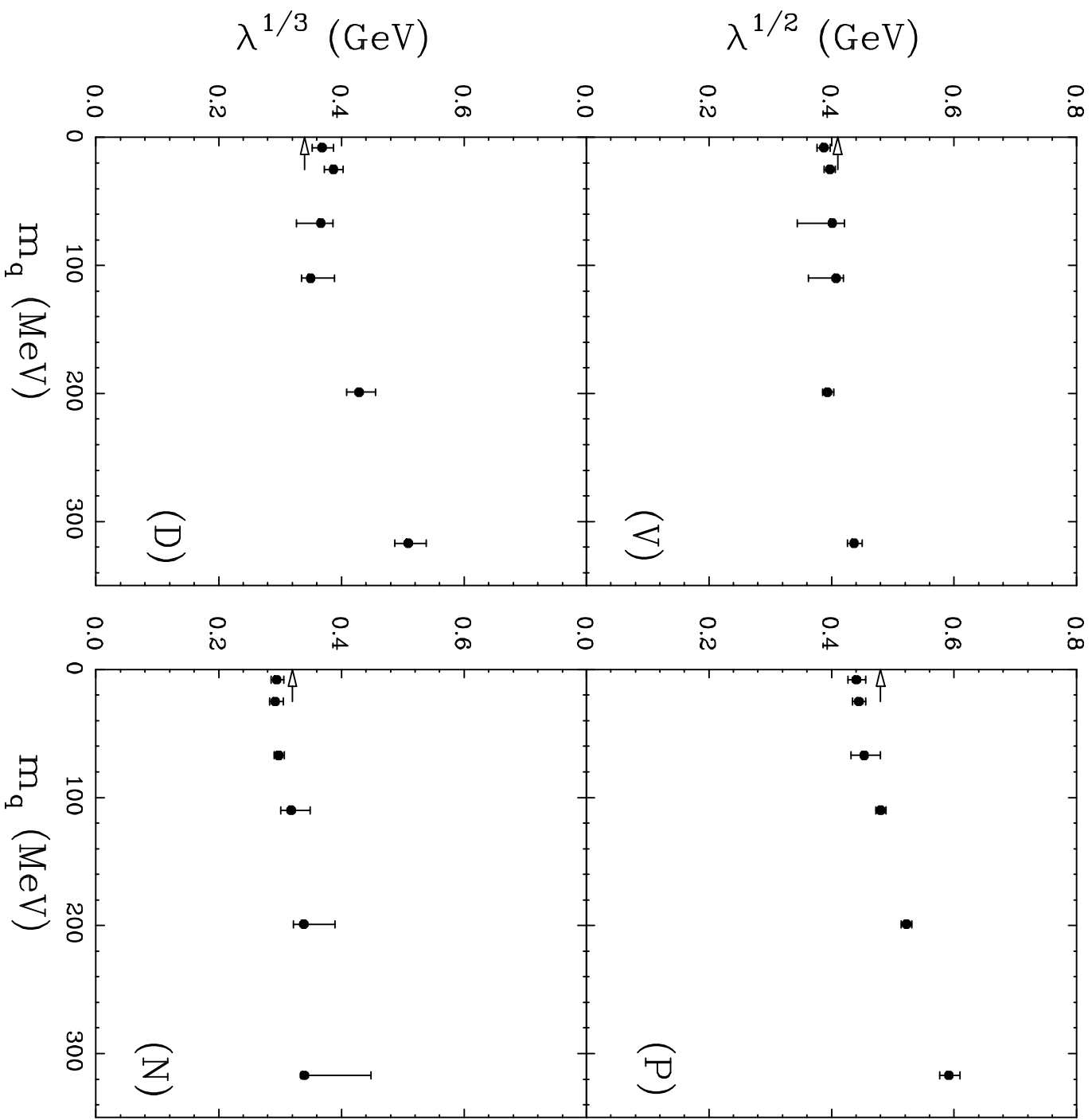
\end{document}